\documentclass[11pt]{article}
\usepackage{amsmath,amsxtra}
\usepackage{amscd}
\usepackage[dvips]{graphicx}
\usepackage{graphicx}
\usepackage{latexsym}
\usepackage{latexsym}
\usepackage{amsmath}
\date{\nonumber}
\begin{document}
\date{}
%  Theorems, Lemmas and the like, should be typeset in italic
\newtheorem{theorem}{Theorem}
\newtheorem{proposition}{Proposition}
\newtheorem{lemma}{Lemma}
\newtheorem{definition}{Definition}

\renewcommand{\theequation}
{\arabic{section}.\arabic{equation}}
%%%%% END DOCUMENT SPECIFIC DEFINITIONS
%\renewcommand{\square}{\hfill$\Box$\vspace{2ex}}
%\renewcommand{\Theta}{\Ta}

\renewcommand{\theequation}
{\arabic{section}.\arabic{equation}}
\title{Camassa-Holm  and M-CIV  equations with self-consistent sources: geometry and peakon solutions}
\author{Gulmira Yergaliyeva\footnote{Email: gyergaliyeva@gmail.com},  \,           Tolkynay Myrzakul\footnote{Email: trmyrzakul@gmail.com}, \, Gulgassyl  Nugmanova\footnote{Email: nugmanovagn@gmail.com},  \,\\ Kuralay  Yesmakhanova\footnote{Email: kryesmakhanova@gmail.com} \, and Ratbay Myrzakulov\footnote{Email: rmyrzakulov@gmail.com}
\\
Eurasian International Center for Theoretical Physics, \\
Eurasian National University, Nur-Sultan, 010008, Kazakhstan
} \maketitle
\begin{abstract}
In this paper, we study one of  generalized Heisenberg ferromagnet  equations with self-consistent sources, namely, the so-called M-CIV equation with self-consistent sources (M-CIVESCS).  The Lax representation of the M-CIVESCS is presented. We have shown that the M-CIVESCS and the CH equation with self-consistent sources (CHESCS) is geometrically 
equivalent each to other.  The gauge equivalence between these equations is proved. Soliton (peakon) and pseudo-spherical surfaces induced by these  equations are considered. The one  peakon solution of the M-CIVESCS is presented.
\end{abstract}

\vskip .3cm {\bf KEYWORDS:} Camassa-Holm equation with
self-consistent sources; Heisenberg ferromagnet  equation; Heisenberg ferromagnet  equation with
self-consistent sources; Heisenberg ferromagnet  equation with
self-consistent potentials; Lax representation; conservation laws;
peakon; soliton. \vskip .1cm
\date{}
%\tableofcontents
\section{Introduction}
Camassa-Holm  equation (CHE) has the form
\begin{equation}
u_t+2\omega u_x-u_{xxt}+3uu_x=2u_xu_{xx}+uu_{xxx},
\end{equation}
where $u=u(x,t)$ is the fluid velocity in the $x$ direction and 
 $\omega=const$ is related to the critical shallow water wave
speed. This equation has   several equivalent forms, for example, the following ones 
\begin{eqnarray}
q_t+2u_xq+uq_x&=&0, \label{1.2}\\
q-u+u_{xx}-\omega&=&0, \label{1.3}
\end{eqnarray}
or
\begin{eqnarray}
\kappa_t+ (u\kappa)_x&=&0, \label{1.4}\\
u-u_{xx}+\omega-\nu\kappa^{2}&=&0. \label{1.5}
\end{eqnarray}
The CHE  was
implicitly contained in the class of multi-Hamiltonian system
introduced by Fuchssteiner and Fokas \cite{1}. It  explicitly derived
as a shallow water wave equation by Camassa and Holm \cite{2} and after their  works, the CHE and its different  modifications  have been studied from
many kinds of views \cite{0811.2552}-\cite{ivanov}. 
The CHE  shares most of
the properties of the integrable system of KdV type and 
possesses Lax representation (LR),  the bi-Hamiltonian structure, smooth solitary wave solutions (as $\omega>0$). For example, the solitary wave  solutions become piecewise smooth and
have cusps at their peaks when
$\omega\rightarrow 0$.  When $\omega=0$, these kind of solutions are weak
solutions  and are called "peakons".

Integrable Heisenberg ferromagnet type equations play important role in modern physics and mathematics (see, e.g., Refs.  \cite{assem1}-\cite{G4}). Recently, integrable generalized Heisenberg ferromagnet equations which are equivalent to Camassa-Holm type equations were presented (see, e.g., refs. \cite{assem1}-\cite{bayan2}). In particular, it is shown  that the CHE is (geometrically and gauge) equivalent to the following Myrzakulov-CIV (M-CIV) equation \cite{assem1}-\cite{1907.10910}
\begin{eqnarray}
 A_{xt}+(uA_{x})_{x}+(uA_{x}^{2}+0.5\{A_{x},A_{t}\})A-\frac{1}{4\beta^{2}}[A,A_{x}]&=&0,\label{1.6}\\
tr(A_{x}^{2})+8\beta^{2}(u-u_{xx})&=&0 \label{1.7}
\end{eqnarray}
or
\begin{eqnarray}
(A_{t}+uA_{x})_{x}+(u_{xx}-u_{x}-2u)A-\frac{1}{4\beta^{2}}[A,A_{x}]&=&0,\label{1.8}\\
tr(A_{x}^{2})+8\beta^{2}(u-u_{xx})&=&0. \label{1.9}
\end{eqnarray}
Here
\begin{eqnarray}
A&=&\left(\begin{array} {cc} A_{3}& A^{-} \\ 
A^{+}&-A_{3}\end{array}\right), \quad A^{\pm}=A_{1}\pm iA_{2}, \quad A^{2}=I, \quad {\bf A}=(A_{1}, A_{2}, A_{3}), \quad {\bf A}^{2}=1, \label{1.10}\\
A_{x}^{2}&=&-4\beta^{2}qI, \quad \{A_{t},A_{x}\}=[(8\beta^{2}u-4)q-2(u_{x}+u_{xx})]I. \label{1.11}
\end{eqnarray}
There are many integrable generalizations of the CHE. One of such generalizations is the CHE with self-consistent sources (CHESCS) \cite{0811.2552}. Such type integrable  equations with self-consistent sources  have
attracted much attention in recent years \cite{r5}. They have  important
applications  in many branches  of physics. For example, the nonlinear
Schr$\ddot{o}$dinger equation with self-consistent sources describes  the nonlinear interaction of an electrostatic
high-frequency wave with the ion acoustic wave in a two component
homogeneous plasma. Another example is the KdV
equation with self-consistent sources. It  describes the interaction of
long and short capillary-gravity waves. The famous KP equation with self-consistent
sources represents  the nonlinear interaction of a long wave with a short wave
packet propagating on the $x$-$y$ plane at some angle to each other. 
In this paper, we would like to study the M-CIV equation with self-consistent sources (M-CIVESCS) and its  relation with  the CHESCS.

This paper is organized as follows. In Section 2, we present  the M-CIVESCS and its Lax representation. In Section 3, the integrable motion of curves induced by  the M-CIVESCS are constructed.  The gauge equivalence between the M-CIVESCS and CHESCS is established in Section 4. The peakon (soliton) and pseudo-spherical  surfaces induced by the M-CIVESCS and the CHESCS are presented in Section 5 and in Section 6, respectively. In Section 7,  the formulas of the one 
peakon solution of the M-CIVESCS is presented.
The M-CIVESCS and the CHESCS with $N$ self-consistent sources is given in Section 8.  In Section 9,
the conclusion is presented.

%%%%%%%%%%%%%%%%%%%%%%%%%%%%%%%%%%%%%%%%%%%%%%%%%%%%%%%
\section{M-CIV equation with self-consistent sources}
%%%%%%%%%%%%%%%%%%%%%%%%%%%%%%%%%%%%%%%%%%%%%%%%%%%%%%%%

One of examples of the peakon spin systems is the following Myrzakulov-CIV  equation with self-consistent sources (M-CIVESCS) 
\begin{eqnarray}
 [A,A_{xt}] +(u[A,A_{x}])_{x}-\frac{1}{\beta^{2}}A_{x}-\alpha(\omega[A,A_{x}])_{x}+\frac{2\alpha^{2}\chi}{\beta^{2}+\alpha}A_{x}+\frac{2\alpha\beta^{2}\chi_{x}}{\beta^{2}+\alpha}A &=&0, \label{2.12}\\
\psi_{1x}-(\frac{\alpha}{4\beta}-\frac{1}{4})[(A^{+}_{x}A^{-}-A^{+}A^{-}_{x})\psi_{1}+2(A^{-}_{x}A_{3}-A^{-}A_{3x})\psi_{2}]&=&0, \label{2.13}\\
\psi_{2x}-(\frac{\alpha}{4\beta}-\frac{1}{4})[2(A^{+}A_{3x}-A^{+}_{x}A_{3})\psi_{1}+(A^{+}A^{-}_{x}-A^{+}_{x}A^{-})\psi_{2}]&=&0.\label{2.14}
\end{eqnarray}
This equation can be written in the following equivalent form
\begin{eqnarray}
 A_{xt}+uA_{xx}+(uA_{x}^{2}+0.5\{A_{x},A_{t}\})A +u_{1}A_{x}+u_{3}I+u_{2}[A,A_{x}] &=&0, \label{2.15}\\
\psi_{1x}-(\frac{\alpha}{4\beta}-\frac{1}{4})[(A^{+}_{x}A^{-}-A^{+}A^{-}_{x})\psi_{1}+2(A^{-}_{x}A_{3}-A^{-}A_{3x})\psi_{2}]&=&0, \label{2.16}\\
\psi_{2x}-(\frac{\alpha}{4\beta}-\frac{1}{4})[2(A^{+}A_{3x}-A^{+}_{x}A_{3})\psi_{1}+(A^{+}A^{-}_{x}-A^{+}_{x}A^{-})\psi_{2}]&=&0, \label{2.17}
\end{eqnarray}
where $\omega=\phi_{1}^{2}$, $\chi=\omega_{x}+\omega$,  $\alpha, \beta=consts$ and
\begin{eqnarray}
u_{1}&=&u-\alpha\omega, \quad u_{2}=\frac{\alpha^{2}\chi}{\beta^{2}+\alpha}-
\frac{1}{4\beta^{2}},\quad u_{3}=\frac{\alpha\beta^{2}\chi_{x}}{\beta^{2}+\alpha}, \label{2.18}\\ 
A&=&\left(\begin{array} {cc} A_{3}& A^{-} \\ 
A^{+}&-A_{3}\end{array}\right), \quad A^{\pm}=A_{1}\pm iA_{2}, \quad A^{2}=I, \quad {\bf A}=(A_{1}, A_{2}, A_{3}), \quad {\bf A}^{2}=1.\label{2.19}
\end{eqnarray}
The LR of the M-CIVESCS   reads as
\begin{eqnarray}
\Psi_{x}&=&U_{1}\Psi, \label{2.20}\\
 \Psi_{t}&=&V_{1}\Psi.\label{2.21}
\end{eqnarray}
Here
\begin{eqnarray}
U_{1}&=&\left(\frac{\lambda}{4\beta}-\frac{1}{4}\right)[A,A_{x}], \label{2.22}\\
V_{1}&=&\left(\frac{1}{4\beta^{2}}-\frac{1}{4\lambda^{2}}\right)A+\left[\frac{1}{8\beta\lambda}-\frac{1}{8\beta^{2}}-\left(\frac{\lambda}{4\beta}-\frac{1}{4}\right)u\right][A,A_{x}]+\left(\frac{1}{2\lambda}-\frac{1}{2\beta}\right)Z,\label{2.23}
\end{eqnarray}
where
\begin{eqnarray}
Z=\frac{1}{2}\left[\frac{u_{x}+u_{xx}}{\beta}-\frac{\alpha\beta\chi_{x}}{\beta^{2}+\alpha}\right]^{-1}\left[A,A_{t}-\left(\frac{1}{2\beta}-u+\frac{\alpha\beta\omega}{\beta^{2}+\alpha}\right)A_{x}\right].\label{2.24}
\end{eqnarray}

%%%%%%%%%%%%%%%%%%%%%%%%%%%%%%%%%%%%%%%%%%%%%%%%%%%%%%%%%%
\section{Integrable motion of space curves induced by the M-CIVESCS}
%%%%%%%%%%%%%%%%%%%%%%%%%%%%%%%%%%%%%%%%%%%%%%%%%%%%%%%%%%%%%%%

In this section, we  consider  the integrable motion of space curves induced by the  M-CIVESCS.   As usual, let us consider a smooth space curve ${\bf \gamma} (x,t): [0,X] \times [0, T] \rightarrow R^{3}$ in $R^{3}$. Let  $x$ is the arc length of the curve at each time $t$.   The   corresponding Frenet-Serret equation and its temporal counterpart look like 
\begin{eqnarray}
\left ( \begin{array}{ccc}
{\bf  e}_{1} \\
{\bf  e}_{2} \\
{\bf  e}_{3}
\end{array} \right)_{x} = C
\left ( \begin{array}{ccc}
{\bf  e}_{1} \\
{\bf  e}_{2} \\
{\bf  e}_{3}
\end{array} \right),\quad
\left ( \begin{array}{ccc}
{\bf  e}_{1} \\
{\bf  e}_{2} \\
{\bf  e}_{3}
\end{array} \right)_{t} = G
\left ( \begin{array}{ccc}
{\bf  e}_{1} \\
{\bf  e}_{2} \\
{\bf  e}_{3}
\end{array} \right), \label{3.25} 
\end{eqnarray}
where ${\bf e}_{j}$ are the   unit tangent vector $(j=1)$,  principal normal vector $(j=2)$ and binormal vector $(j=3)$ which given by ${\bf e}_{1}={\bf \gamma}_{x}, \quad {\bf e}_{2}=\frac{{\bf \gamma}_{xx}}{|{\bf \gamma}_{xx}|}, \quad {\bf e}_{3}={\bf e}_{1}\wedge {\bf e}_{2}, $ 
respectively.
Here
\begin{eqnarray}
C=
\left ( \begin{array}{ccc}
0   & \kappa_{1}     & \kappa_{2}  \\
-\kappa_{1}  & 0     & \tau  \\
-\kappa_{2}    & -\tau & 0
\end{array} \right),\quad
G=
\left ( \begin{array}{ccc}
0       & \omega_{3}  & \omega_{2} \\
-\omega_{3} & 0      & \omega_{1} \\
-\omega_{2}  & -\omega_{1} & 0
\end{array} \right),\label{3.26} 
\end{eqnarray}
where $\tau$,  $\kappa_{1}, \kappa_{2}$ are the  "torsion",  "geodesic curvature" and  "normal curvature" of the curve, respectively; $\omega_{j}$ are some  functions.  The compatibility condition of the equations (\ref{3.25}) reads  as
\begin{eqnarray}
C_t - G_x + [C, G] = 0\label{3.27} 
\end{eqnarray}
or in elements   
 \begin{eqnarray}
\kappa_{1t}- \omega_{3x} -\kappa_{2}\omega_{1}+ \tau \omega_2&=&0, \label{53} \\ 
\kappa_{2t}- \omega_{2x} +\kappa_{1}\omega_{1}- \tau \omega_3&=&0, \label{54} \\
\tau_{t}  -    \omega_{1x} - \kappa_{1}\omega_2+\kappa_{2}\omega_{3}&=&0.  \label{55} \end{eqnarray}
We now assume ${\bf A}\equiv {\bf e}_{1}$. Let take place the following expressions    
\begin{eqnarray}
\kappa_{1}=i, \quad \kappa_{2}=\lambda(q-1), \quad \tau=-i\lambda(q+1), \label{57} 
\end{eqnarray}
where $q=0.5\lambda(\kappa_{2}+i\tau)$.  Then we have 
\begin{eqnarray}
\omega_{1} & = &i[(0.5\lambda^{-1}-\lambda u)(q+1)+0.5\lambda^{-2}(u_{x}+u_{xx})],\label{58}\\ 
\omega_{2}&=& [(0.5\lambda^{-1}-\lambda u)(q+1)+0.5\lambda^{-1}(u_{x}+u_{xx})], \label{59} \\
\omega_{3} & = &i[0.5\lambda^{-2}-u-u_{x}].      \label{60}
\end{eqnarray}
Now Eqs.(\ref{53})-(\ref{55}) give us the following equations for $q, u, \phi_{1}$:
\begin{eqnarray}
q_t+2qu_x+uq_x-\omega_{x}+\omega_{xxx}&=&0,\label{3.35}\\
\phi_{1xx}-(\alpha^{2}q+\frac{1}{4})\phi_{1}&=&0 \label{3.36}
\end{eqnarray}
or
\begin{eqnarray}
q_t+2qu_x+uq_x-[(\phi_{1}^2)_x-(\phi_{1}^2)_{xxx}]&=&0,\\
\phi_{1xx}-(\alpha^{2}q+\frac{1}{4})\phi_{1}&=&0. \label{90}
\end{eqnarray}
It is nothing but the CHESCS \cite{0811.2552}-\cite{Iroda}.  
So, we have  proved  that   the M-CIVESCS   is the  Lakshmanan (geometrical)  equivalent to the CHESCS.

%%%%%%%%%%%%%%%%%%%%%%%%%%%%%%%%%%%%%%%%%%%%%%%%%%%%%%%%%%%%%%%%%
%%%%%%%%%%%%%%%%%%%%%%%%%%%%%%%%%%%%%%%%%%%%
 \section{Gauge equivalence between the M-CIVESCS and the CHESCS}
%%%%%%%%%%%%%%%%%%%%%%%%%%%%%%%%%%%%%%%%%%%%%%%%
Above, we have shown that the M-CIVESCS and the CHESCS are the geometrical (Lakshmanan) equivalent each to other. Now  we consider the possible gauge equivalence between these equations \cite{gulmira1}.  First, we note that from the results of the previous section and from the isomorphism $so(3)\approx su(2)$ follow the  LR for the CHESCS of the form \cite{0811.2552}
\begin{eqnarray}
\Phi_{x}&=&U_{2}\Phi,\label{4.39}\\
\Phi_{t}&=&V_{2}\Phi, \label{4.40}
\end{eqnarray}
where
\begin{eqnarray}
\Phi=\left( \begin{array}{c}
 \phi_{1}     \\
\phi_{2}
\end{array} \right), \quad U_{2}=\left ( \begin{array}{cc}
 -0.5  &  \lambda   \\
\lambda q   & 0.5
\end{array} \right), \quad
V_{2}=V_{20}+V_{2}^{'}.
\end{eqnarray}
Here
\begin{eqnarray}
V_{20}&=&\left ( \begin{array}{cc}
 \frac{u_{x}+u_{xx}}{2}-\frac{1}{4\lambda^{2}}  &  \frac{1}{2\lambda} - u\lambda  \\
\frac{u_{x}+u_{xx}+q}{2\lambda}-uq\lambda   & \frac{1}{4\lambda^{2}}-\frac{u_{x}+u_{xx}}{2}
\end{array} \right), \\
V_{2}^{'}&=&-\frac{\alpha\lambda^{2}\chi}{2(\lambda^{2}+\alpha)}\sigma_{3}+\frac{\alpha\lambda^{3}\omega}{\lambda^{2}+\alpha}\left ( \begin{array}{cc}
0    & 1 \\
q    & 0
\end{array} \right)-\frac{\alpha\lambda\chi_{x}}{2(\lambda^{2}+\alpha)}\Sigma
, \quad 
\Sigma=\left( \begin{array}{cc}
 0  &  0   \\
1   & 0
\end{array} \right). 
\end{eqnarray}
The compatibility condition $\Phi_{xt}=\Phi_{tx}$ given by 
\begin{eqnarray}
U_{2t}-V_{2x}+[U_{2},V_{2}]=0 \label{9} 
\end{eqnarray}
is equivalent to the CHESCS (\ref{3.35})-(\ref{3.36}).
Consider the transformation 
   $\Psi=g^{-1}\Phi$, where $\Psi$ is the solution of the equations (\ref{2.20})-(\ref{2.21}) and $g=\Phi|_{\lambda=\beta}$. Then the Lax pairs of the M-CIVESCS and CHESCS is related by the following equations
	\begin{eqnarray}
U_{1}=g^{-1}U_{2}g-g^{-1}g_{x}, \quad V_{1}=g^{-1}V_{2}g-g^{-1}g_{t}.
\end{eqnarray}
Note that these LR can be rewritten in the  equivalent scalar forms. For example, the equivalent scalar form of the LR for the CHESCS is given by \cite{0811.2552}
\begin{eqnarray}
\phi_{1xx}&=&(\lambda^{2} q+\frac{1}{4})\phi_{1},\\
\phi_{1t}&=&\left(\frac{1}{2\lambda^{2}}-u+\frac{\alpha\lambda^{2}\omega}{\lambda^{2}+\alpha}\right)\phi_{1x}+\left(\frac{1}{2}u_x-\frac{\alpha\lambda^{2}\omega_{x}}{2(\lambda^{2}+\alpha)}\right)\phi_{1}.
\end{eqnarray}
The compatibility condition $\phi_{1xxt}=\phi_{1txx}$ is equivalent to the CHESCS (\ref{3.35})-(\ref{3.36}).
Finally we present the following important relation between the solutions of the M-CIVESCS and the CHESCS:
\begin{eqnarray}
tr(A_{x}^{2})=-8\beta^{2}q=-8\beta^{2}(u-u_{xx})
\end{eqnarray}
or
\begin{eqnarray}
{\bf A}_{x}^{2}=-4\beta^{2}q=-4\beta^{2}(u-u_{xx}).
\end{eqnarray}

%%%%%%%%%%%%%%%%%%%%%%%%%%%%%%%%%%%%%%%%%%%%%%%%%%%%%%%%
\section{Peakon (soliton) surfaces}
%%%%%%%%%%%%%%%%%%%%%%%%%%%%%%%%%%%%%%%%%%%%%%%%%%%%%
%%%%%%%%%%%%%%%%%%%%%%%%%%%%%%%%%%%%%%%%%%%%%%
\subsection{Peakon surfaces corresponding to the CHESCS}
%%%%%%%%%%%%%%%%%%%%%%%%%%%%%%%%%%%%%%%%%%%%%%%%%%%%%%%%%

As well-known, the Sym-Tafel formula gives an interesting connection between the classical geometry of manifolds  immersed in $R^{n}$ and the integrable systems.  Using the Sym-Tafel formula, here we want to construct the soliton (peakon) surface induced by the CHESCS and the M-CIVESCS. To this end, let us consider a $\lambda$-family of parametric surfaces given by the matrix $r=r(x,t,\lambda)$. According to the Sym-Tafel formula, this matrix defines as
\begin{eqnarray}
r=\Phi^{-1}\Phi_{\lambda}, \label{5.50}
\end{eqnarray}
where $\Phi$ is the solution of the equations (\ref{4.39})-(\ref{4.40}). We have
\begin{eqnarray}
r_{x}=\Phi^{-1}U_{2\lambda}\Phi, \quad r_{t}=\Phi^{-1}V_{2\lambda}\Phi.\label{0}
\end{eqnarray}
Then  the components of the metric tensor define as
\begin{eqnarray}
g_{ij}={\bf r}_{,i}\cdot{\bf r}_{,j}, \label{5.52}
\end{eqnarray}
where ${\bf r}=(r_{1},r_{2},r_{3})$ is the position vector, ${\bf r}_{,1}\equiv{\bf r}_{x}, \quad {\bf r}_{,2}\equiv{\bf r}_{t}$. Then for example, the  first fundamental form of the soliton surface is given by
\begin{eqnarray}
I=g_{ij}dx^{i}dx^{j}={\bf r}_{x}^{2}dx^{2}+2{\bf r}_{x}{\bf r}_{t}dxdt+{\bf r}_{t}^{2}dt^{2}, \label{0}
\end{eqnarray}
where
\begin{eqnarray}
{\bf r}_{x}^{2}=\frac{1}{2}tr(r_{x}^{2}), \quad {\bf r}_{x}{\bf r}_{t}=\frac{1}{2}tr(r_{x}r_{t}), \quad {\bf r}_{t}^{2}=\frac{1}{2}tr(r_{t}^{2}).\label{0}
\end{eqnarray}
For example,
\begin{eqnarray}
g_{11}={\bf r}_{x}^{2}=q. \label{0}
\end{eqnarray}
Similarly, we can construct the second fundamental form in the standard way (see, e.g., \cite{jan}). 
%%%%%%%%%%%%%%%%%%%%%%%%%%%%%%%%%%%%%%%%%%%%%%%%%%%%%%%%%%%
\subsection{Peakon surfaces corresponding to the M-CIVESCS}
%%%%%%%%%%%%%%%%%%%%%%%%%%%%%%%%%%%%%%%%%%%%%%%%
Let us now we present the main elements of the soliton (peakon)  surfaces induced by the M-CIVESCS. In this case, the matrix  $r$ is given by
\begin{eqnarray}
r=\Psi^{-1}\Psi_{\lambda}. \label{0}
\end{eqnarray}
Here $\Psi$ is the solution of the equations (\ref{2.20})-(\ref{2.21}). These equations give
\begin{eqnarray}
r_{x}=\Psi^{-1}U_{1\lambda}\Psi, \quad r_{t}=\Psi^{-1}V_{1\lambda}\Psi.\label{0}
\end{eqnarray}
Then  the components of the metric tensor is given by the formula (\ref{5.52}). 
For example,
\begin{eqnarray}
g_{11}=-\frac{1}{8\beta^{2}}tr(A_{x}^{2}). \label{5.55}
\end{eqnarray}
Now it is not difficult to construct the first and  second fundamental forms of  the peakon (soliton) surfaces that we can do  in the standard way (see, e.g., \cite{jan}). 

\section{Pseudo-spherical surfaces}
%%%%%%%%%%%%%%%%%%%%%%%%%%%%%%%%%%%%%%%%%%%%%%%%%%
In this section, we briefly present the main facts  on the pseudo-spherical surfaces induced by the M-CIVESCS  and the CHESCS. %%%%%%%%%%%%%%%%%%%%%%%%%%%%%%%%%%%%%%%%%%%
\subsection{Pseudo-spherical  surfaces related with  the CHESCS}
%%%%%%%%%%%%%%%%%%%%%%%%%%%%%%%%%%%%%%%%%%%%%%
Consider the following linear problem
\begin{eqnarray}
d\Phi=Y_{2}\Phi,
\end{eqnarray}
where
\begin{eqnarray}
Y_{2}=\frac{1}{2}\left( \begin{array}{cc}
 \omega^{2}  &  \omega^{1}-\omega^{3}  \\
\omega^{1}+\omega^{3}   & -\omega^{2}
\end{array} \right)= U_{2}dx+V_{2}dt. 
\end{eqnarray}
The integrable condition of the 1-form $Y_{2}$ is given by
\begin{eqnarray}
d^{2}Y_{2}=(U_{2t}-V_{2x}+[U_{2},V_{2}])dt\wedge dx=0. 
\end{eqnarray}
This means  that the one-forms $\omega_{j}$ satisfy the following structure equations 
\begin{eqnarray}
d\omega^{1}&=&\omega^{3}\wedge\omega^{2}, \\
d\omega^{2}&=&\omega^{1}\wedge\omega^{3}, \\
d\omega^{3}&=&\omega^{1}\wedge\omega^{2}.   
\end{eqnarray}
These equations are equivalent  to  the CHESCS. 
%%%%%%%%%%%%%%%%%%%%%%%%%%%%%%%%%%%%%%%%%%%
\subsection{Pseudo-spherical  surfaces related with  the M-CIVESCS}
%%%%%%%%%%%%%%%%%%%%%%%%%%%%%%%%%%%%%%%%%%%%%%
We now return to the M-CIVESCS. To construct the pseudo-spherical surfaces induced by this equation, let us  consider the linear problem
\begin{eqnarray}
d\Psi=Y_{1}\Psi,
\end{eqnarray}
where the 1-form $Y_{1}$ reads as
\begin{eqnarray}
Y_{1}=\frac{1}{2}\left( \begin{array}{cc}
 \sigma^{2}  &  \sigma^{1}-\sigma^{3}  \\
\sigma^{1}+\sigma^{3}   & -\sigma^{2}
\end{array} \right)= U_{1}dx+V_{1}dt. 
\end{eqnarray}
As in the previous subsection, we consider the integrable condition of the  1-form $Y_{1}$:
\begin{eqnarray}
d^{2}Y_{1}=(U_{1t}-V_{1x}+[U_{1},V_{1}])dt\wedge dx=0 
\end{eqnarray}
which  in components takes the form
\begin{eqnarray}
d\sigma^{1}&=&\sigma^{3}\wedge\sigma^{2}, \\
d\sigma^{2}&=&\sigma^{1}\wedge\sigma^{3}, \\
d\sigma^{3}&=&\sigma^{1}\wedge\sigma^{2}.   
\end{eqnarray}
These equations is equivalent to the M-CIVESCS. Thus the M-CIVESCS and the CHESCS describe some kind pseudo-spherical surfaces as their analogies without sources (see, e.g., refs. \cite{reyes1}-\cite{reyes2}).

%%%%%%%%%%%%%%%%%%%%%%%%%%%%%%%%%%%%%%%%%%%%%%%%%%%%%%%%%%
%%%%%%%%%%%%%%%%%%%%%%%%%%%%%%%%%%%%%%%%%%%%%%%
\section{One peakon solution of the M-CIV ESCS}
%%%%%%%%%%%%%%%%%%%%%%%%%%%%%%%%%%%%%%%%%%%%
As the integrable equation, the M-CIVESCS has all ingredients of integrable systems like LR, conservation laws, bi-Hamiltonian structure, soliton solutions and so on. In particular, it admits the peakon solutions. Here let us present a one peakon solution of the M-CIVESCS. To construct this 1-peakon solution, we use the corresponding 1-peakon solution of the CHESCS \cite{0811.2552}. The 1-peakon solution of the M-CIVESCS has the form
\begin{eqnarray}
A^{+}&=&\frac{2g_{1}g_{2}}{|g_{1}|^{2}+|g_{2}|^{2}}, \quad A_{3}=\frac{|g_{1}|^{2}-|g_{2}|^{2}}{|g_{1}|^{2}+|g_{2}|^{2}}, \\
\psi_{1}&=&\frac{\bar{g}_{1}\phi_{1}+\bar{g}_{2}\phi_{2}}{|g_{1}|^{2}+|g_{2}|^{2}}, \quad \psi_{2}=\frac{-g_{2}\phi_{1}+g_{1}\phi_{2}}{|g_{1}|^{2}+|g_{2}|^{2}}.  
\end{eqnarray}
Here
\begin{eqnarray}
\phi_{2}=\lambda^{-1}(\phi_{1x}+0.5\phi_{1}),\quad g_{j}=\phi_{j}|_{\lambda=\beta},
\end{eqnarray}
where
\begin{eqnarray}
\phi_{1}=\sqrt{\sigma'(t)c}e^{-\frac{1}{2}|x-ct+\sigma(t)|}, \quad u=ce^{-|x-ct+\sigma(t)|}
\end{eqnarray}
is the 1-peakon solution of the CHESCS \cite{0811.2552}.
%%%%%%%%%%%%%%%%%%%%%%%%%%%%%%%%%%%%%%%%%%%%%%%%%%%%%
\section{Integrable self-consistent $N$  sources case}
%%%%%%%%%%%%%%%%%%%%%%%%%%%%%%%%%%%%%%%%
In the previous sections,  we have considered the M-CIVESCS and the CHESCS with the one self-consistent source. In this section, we present, in short form,   integrable generalizations of these  equations with  $N$ self-consistent sources. 
\subsection{M-CIV equation with self-consistent $N$-sources}
The M-CIVE with $N$ self-consistent sources has the form
 \begin{eqnarray}
 [A,A_{xt}+(uA_{x})_{x}]-\frac{1}{\beta^{2}}A_{x}-\sum_{j=1}^{N}\left(\lambda_{j}(\omega_{j}[A,A_{x}])_{x}+\frac{2\lambda_{j}^{2}\chi_{j}}{\beta^{2}+\lambda_{j}}A_{x}+\frac{2\lambda_{j}\beta^{2}\chi_{jx}}{\beta^{2}+\lambda_{j}}A \right)&=&0, \label{21}\\
\psi_{1jx}-(\frac{\lambda_{j}}{4\beta}-\frac{1}{4})[(A^{+}_{x}A^{-}-A^{+}A^{-}_{x})\psi_{1j}+2(A^{-}_{x}A_{3}-A^{-}A_{3x})\psi_{2j}]&=&0, \label{43}\\
\psi_{2jx}-(\frac{\lambda_{j}}{4\beta}-\frac{1}{4})[2(A^{+}A_{3x}-A^{+}_{x}A_{3})\psi_{1j}+(A^{+}A^{-}_{x}-A^{+}_{x}A^{-})\psi_{2j}]&=&0, \label{43}
\end{eqnarray}
where $\omega_{j}=\phi_{1j}^{2}$, $\chi_{j}=\omega_{jx}+\omega_{j}$,  $\alpha, \beta=consts$ and
\begin{eqnarray}
A=\left(\begin{array} {cc} A_{3}& A^{-} \\ 
A^{+}&-A_{3}\end{array}\right), \quad A^{\pm}=A_{1}\pm iA_{2}, \quad A^{2}=I, \quad {\bf A}=(A_{1}, A_{2}, A_{3}), \quad {\bf A}^{2}=1.
\end{eqnarray}
The LR of the M-CIVESCS   reads as
\begin{eqnarray}
\Psi_{x}&=&U_{1}\Psi,\\
 \Psi_{t}&=&V_{1}\Psi.
\end{eqnarray}
Here
\begin{eqnarray}
U_{1}&=&\left(\frac{\lambda}{4\beta}-\frac{1}{4}\right)[A,A_{x}],\\
V_{1}&=&\left(\frac{1}{4\beta^{2}}-\frac{1}{4\lambda^{2}}\right)A+\left[\frac{1}{8\beta\lambda}-\frac{1}{8\beta^{2}}-\left(\frac{\lambda}{4\beta}-\frac{1}{4}\right)u\right][A,A_{x}]+\left(\frac{1}{2\lambda}-\frac{1}{2\beta}\right)\sum_{j=1}^{N}Z_{j},
\end{eqnarray}
where
\begin{eqnarray}
Z_{j}=\frac{1}{2}\left[\frac{u_{x}+u_{xx}}{\beta}-\frac{\lambda_{j}\beta\chi_{jx}}{\beta^{2}+\lambda_{j}}\right]^{-1}\left[A,A_{t}-\left(\frac{1}{2\beta}-u+\frac{\lambda_{j}\beta\omega_{j}}{\beta^{2}+\lambda_{j}}\right)A_{x}\right].
\end{eqnarray}
%%%%%%%%%%%%%%%%%%%%%%%%%%%%%%%%%%%%%%%%%%%%%%%
\subsection{CHE with self-consistent  $N$-sources}
%%%%%%%%%%%%%%%%%%%%%%%%%%%%%%%%%%%%%%%%%%%%%%%%%%%
The CHE with $N$ self-consistent sources has the form \cite{0811.2552}
\begin{eqnarray}
q_t+2qu_x+uq_x-\sum_{j=1}^N[(\varphi_j^2)_x-(\varphi_j^2)_{xxx}]&=&0,\\
\varphi_{j,xx}-(\lambda_j q+\frac{1}{4})\varphi_j&=&0.
\end{eqnarray}
Its LR reads as \cite{0811.2552}
\begin{eqnarray}
\phi_{xx}&=&(\frac{1}{4}+\lambda q)\phi,\\
\phi_{t}&=&\frac{u_{x}}{2}\phi+(\frac{1}{2\lambda
}-u)\phi_{x}+2 \sum\limits_{j=1}^{N}\frac{\lambda\lambda_{j}
\phi_{j}}{\lambda-\lambda_{j}}(\phi_{jx}\phi-\phi_{j}\phi_{x}),
 \end{eqnarray}
which means that the CHESCS  is Lax integrable \cite{0811.2552}.

\section{Conclusion}
The integrable generalized Heisenberg ferromagnet equation with self-consistent sources, namely, the  M-CIVESCS is investigated. The integrable motion of space curves induced by the M-CIVESCS is constructed. Using this result, the geometrical equivalence between the M-CIVESCS and the CHESCS is established. It is shown that the M-CIVESCS and the CHESCS  is gauge equivalent each to other. The simplest conservation law and the one peakon solution 
are  constructed. The peakon (soliton) surfaces induced by the M-CIVESCS and the CHESCS are presented. 

\section*{Acknowledgements}
This work was supported  by  the Ministry of Edication  and Science of Kazakhstan under
grants 0118РК00935 and 0118РК00693.

\end{document}